\documentclass[a4paper,12pt]{article}
\usepackage{color}
\usepackage{epsfig}
\usepackage{amsfonts}
\usepackage{amsmath}
\usepackage{amssymb, amsbsy}
\usepackage{mathtools}
\usepackage{chngpage}
\usepackage{rotating}
\usepackage{multirow}
\usepackage{setspace}
\usepackage{bbding, pifont}
\usepackage{url}
\usepackage{array}
\usepackage{caption}

\usepackage{tikz}
\usetikzlibrary{arrows}
\usetikzlibrary{decorations.markings}

\DeclareMathOperator{\se}{se}

\DeclareMathOperator{\logit}{logit}

\begin{document}
\title{Contextualizing selection bias in Mendelian randomization: how bad is it likely to be?}
\author{Apostolos Gkatzionis \textsuperscript{1} and Stephen Burgess \textsuperscript{1,2} \thanks{Corresponding author: Dr Stephen Burgess. Address: MRC Biostatistics Unit, Cambridge Institute of Public Health, Robinson Way, Cambridge, CB2 0SR, UK. Telephone: +44 1223 748651. Fax: none. Email: sb452@medschl.cam.ac.uk.} \\ \\ \\
\textsuperscript{1} MRC Biostatistics Unit, University of Cambridge, UK  \vspace{4mm} \\
\textsuperscript{2} Cardiovascular Epidemiology Unit, \\ Department of Public Health and Primary Care, \\ University of Cambridge, UK }
\maketitle


\clearpage

\setstretch{1.65}
\subsection*{Abstract}
\textbf{Background:} Selection bias affects Mendelian randomization investigations when selection into the study sample depends on a collider between the genetic variant and confounders of the risk factor--outcome association. However, the relative importance of selection bias for Mendelian randomization compared to other potential biases is unclear. \\ \\
\textbf{Methods:} We performed an extensive simulation study to assess the impact of selection bias on a typical Mendelian randomization investigation. We considered inverse probability weighting as a potential method for reducing selection bias. Finally, we investigated whether selection bias may explain a recently reported finding that lipoprotein(a) is not a causal risk factor for cardiovascular mortality in individuals with previous coronary heart disease. \\ \\
\textbf{Results:} Selection bias had a severe impact on bias and Type 1 error rates in our simulation study, but only when selection effects were large. For moderate effects of the risk factor on selection, bias was generally small and Type 1 error rate inflation was not considerable. Inverse probability weighting ameliorated bias when the selection model was correctly specified, but increased bias when selection bias was moderate and the model was misspecified. In the example of lipoprotein(a), strong genetic associations and strong confounder effects on selection mean the reported null effect on cardiovascular mortality could plausibly be explained by selection bias. \\ \\
\textbf{Conclusions:} Selection bias can adversely affect Mendelian randomization investigations, but its impact is likely to be less than other biases. Selection bias is substantial when the effects of the risk factor and confounders on selection are particularly large. \\ \\
\textbf{Keywords:} instrumental variables, causal inference, selection bias, collider bias, inverse probability weighting. \\

\clearpage

\setlength{\fboxrule}{2pt}
\setlength{\fboxsep}{2pt}
\begin{center}
\framebox[\textwidth][c]{
\parbox{0.9 \textwidth}{\smallskip
\textbf{Key messages:}%
\begin{itemize}
\item In Mendelian randomization experiments, selection bias may arise as a result of collider bias when selection depends on the risk factor and/or the outcome.
\item Selection bias is usually small compared to other types of bias if the effects of the risk factor and/or outcome on selection are weak or moderate. However, it can be a real concern if the selection effects are strong. 
\item Selection bias is increased in the presence of strong confounding, or as the proportion of individuals in the population which are eligible for selection increases. It is also influenced by direct confounder or instrument effects on the selection procedure. It is not affected by instrument strength.
\item Inverse probability weighting can be used to adjust for the bias when selection effects are strong and the underlying probability model is correctly specified. However, if selection bias is weak and the probability model is misspecified, inverse probability weighting may even increase the bias.
\end{itemize}
 }
}
\end{center}

\clearpage
\setstretch{1.5}

\section*{Introduction}
Mendelian randomization is the use of genetic information to assess the existence of a causal relationship between a risk factor and an outcome of interest \cite{daveysmith2003, burgess2015book}. It is the application of instrumental variable analysis in the context of genetic epidemiology, where genetic variants are used as instruments. To be a valid instrumental variable, a genetic variant must be associated with the risk factor in a specific way -- it cannot influence the outcome except via its association with the risk factor, and it cannot be associated with any confounder of the risk factor--outcome association. An association between a valid instrumental variable and outcome is indicative of a causal effect of the risk factor on the outcome \cite{angrist1996, didelez2007}.

This paper discusses selection bias in Mendelian randomization. In general, selection bias arises when individuals included in the study population are not a representative sample of the target population \cite{gail2005}. Selection bias is likely to be present in all epidemiological analyses to some extent. Bias due to non-representative selection usually occurs as an example of collider bias \cite{greenland2003, hernan2004, cole2010}. A collider is a variable that is a common effect of two variables (it is causally downstream of both variables). Collider bias occurs when conditioning on such a variable: even if the two initial variables were unrelated (marginally independent), they will typically become related when conditioning on a collider (conditionally dependent). An example of this is the so-called Berkson's bias \cite{hernan2004}: two diseases A and B that often cause hospitalization may be independent across the population, but they will typically be dependent among hospitalized individuals since being hospitalized and not having disease A means one is more likely to have disease B.

Throughout this paper, we assume that risk factor--outcome confounding is represented by a single variable, referred to as the confounder. Collider bias in Mendelian randomization studies often results in a violation of the instrumental variable assumptions. By assumption, an instrumental variable and the confounder are marginally independent. Conditioning on a collider of the instrumental variable and the confounder would induce an association between the two \cite{swanson2015} and would lead to the instrumental variable becoming invalid. Hence, selection bias can lead to an association between the instrumental variable and the outcome in the absence of a causal effect of the risk factor on the outcome \cite{canan2017}.

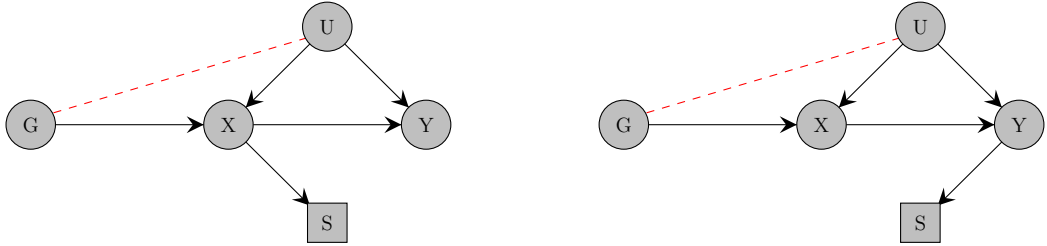
\begin{figure}
\begin{center}
\begin{tikzpicture}[scale=0.65, every node/.style={scale=0.65}]
\draw [decoration={markings,mark=at position 1 with
    {\arrow[scale=2,>=stealth]{>}}},postaction={decorate}] (-9.5, 0) -- (-6.5, 0);
\draw [decoration={markings,mark=at position 1 with
    {\arrow[scale=2,>=stealth]{>}}},postaction={decorate}] (-4.35, 1.65) -- (-5.65, 0.35);
\draw [decoration={markings,mark=at position 1 with
    {\arrow[scale=2,>=stealth]{>}}},postaction={decorate}] (-3.65, 1.65) -- (-2.35, 0.35);
\draw [decoration={markings,mark=at position 1 with
    {\arrow[scale=2,>=stealth]{>}}},postaction={decorate}] (-5.5, 0) -- (-2.5, 0);
    \draw [decoration={markings,mark=at position 1 with
    {\arrow[scale=2,>=stealth]{>}}},postaction={decorate}] (-5.65, -0.35) -- (-4.35, -1.65);
\draw [red, dashed] (-9.85, 0.15) -- (-4.15, 1.85);
\draw[fill = lightgray] (-6, 0) circle (0.5cm);
\draw[fill = lightgray] (-10, 0) circle (0.5cm);
\draw[fill = lightgray] (-2, 0) circle (0.5cm);
\draw[fill = lightgray] (-4, 2) circle (0.5cm);
\draw[fill = lightgray] (-4.4, -2.4) rectangle (-3.6, -1.6);
\node (a) at (-6, 0) {X};
\node (b) at (-10, 0) {G};
\node (c) at (-2, 0) {Y};
\node (d) at (-4, 2) {U};
\node (e) at (-4, -2) {S};

\draw [decoration={markings,mark=at position 1 with
    {\arrow[scale=2,>=stealth]{>}}},postaction={decorate}] (2.5, 0) -- (5.5, 0);
\draw [decoration={markings,mark=at position 1 with
    {\arrow[scale=2,>=stealth]{>}}},postaction={decorate}] (7.65, 1.65) -- (6.35, 0.35);
\draw [decoration={markings,mark=at position 1 with
    {\arrow[scale=2,>=stealth]{>}}},postaction={decorate}] (8.35, 1.65) -- (9.65, 0.35);
\draw [decoration={markings,mark=at position 1 with
    {\arrow[scale=2,>=stealth]{>}}},postaction={decorate}] (6.5, 0) -- (9.5, 0);
    \draw [decoration={markings,mark=at position 1 with
    {\arrow[scale=2,>=stealth]{>}}},postaction={decorate}] (9.65, -0.35) -- (8.35, -1.65);
\draw [red, dashed] (2.15, 0.15) -- (7.85, 1.85);
\draw[fill = lightgray] (6, 0) circle (0.5cm);
\draw[fill = lightgray] (2, 0) circle (0.5cm);
\draw[fill = lightgray] (10, 0) circle (0.5cm);
\draw[fill = lightgray] (8, 2) circle (0.5cm);
\draw[fill = lightgray] (7.6, -2.4) rectangle (8.4, -1.6);
\node (f) at (6, 0) {X};
\node (g) at (2, 0) {G};
\node (h) at (10, 0) {Y};
\node (i) at (8, 2) {U};
\node (j) at (8, -2) {S};

\end{tikzpicture}
\caption{\label{dag} Directed acyclic graphs (DAG) indicating the relationships between an instrumental variable ($G$), exposure ($X$), confounder ($U$) and outcome ($Y$). Selection ($S$) leads to bias if it is a function of the exposure (left panel) or the outcome (right panel), as both exposure and outcome are causally downstream of the genetic variant and confounder, and hence conditioning on selection induces an association between the genetic variant and confounder in both cases.}
\end{center}
\end{figure}

Collider bias in Mendelian randomization can be visualized through causal diagrams. Directed acyclic graphs indicating the relationships between the genetic variant, risk factor, confounder, and outcome are shown in Figure~\ref{dag}. We can see that the risk factor and outcome are both colliders between the genetic variant and the confounder. This means that if selection into the sample population is a function of the risk factor then selection bias will occur (Figure~\ref{dag}, left). The same will occur if selection is a function of the outcome (Figure~\ref{dag}, right), but not if it is a function of the confounder alone, as the confounder is not a collider \cite{hughes2017}.

The possibility that selection bias may undermine instrumental variable analyses, and Mendelian randomization in particular, has long been noted in the literature \cite{nitsch2006}. However, simply saying that selection bias may undermine a Mendelian randomization study is a platitude -- it is a true statement, but not a helpful one. Such unhelpful statements are pervasive in epidemiology papers -- it is common in the discussion of papers analysing observational data to read bald statements highlighting the possibility that findings could have been adversely affected by selection bias, or similar phenomena such as unmeasured confounding and measurement error. It would be more helpful to evaluate to what extent selection bias is likely to influence findings in terms of bias or Type 1 error rate inflation, or to suggest the magnitude of selection bias that would be required for a positive finding to be explained through bias alone \cite{vanderweele2011}.

In this paper, we aim to contextualize to what extent selection bias affects a typical Mendelian randomization investigation. Our hope is that this paper will help investigators make an informed judgement about the relative importance of selection bias in their work compared to other potential sources of bias. We first list some typical scenarios for Mendelian randomization investigations in which selection bias may occur. We then consider simulated datasets that are similar to applied Mendelian randomization investigations, demonstrating the extent of bias and Type 1 error rate inflation that occur due to selecting based on a collider. We show how inverse probability weighting can often help reduce bias and reestablish nominal Type 1 error rates, but this sometimes comes at the cost of increased standard errors for the causal effect estimate. The use of weight trimming to avoid this inflation of standard errors is also discussed. The application of inverse probability weighting is illustrated in an example concerning the effect of lipoprotein(a) on coronary heart disease risk. We also discuss consequences of inverse probability weighting in terms of the population to which the estimate relates.

\section*{Selection bias in practice}
In this section, we provide some examples of Mendelian randomization studies in which selection bias is likely to arise. These are in addition to generic scenarios in which selection bias would occur in any epidemiological analysis, such as the sample population being unrepresentative due to low recruitment rate (for example, the initial recruitment rate for UK Biobank was 10\% \cite{watts2007}) or due to the design of the study (for example, the Million Veteran Program specifically targets US military veterans \cite{gaziano2016}).

\subsection*{Assessing the causal effect of a risk factor on secondary disease or disease progression}
Selection bias could occur when considering a secondary disease outcome, such as disease progression. For example, a recent Mendelian randomization investigation considered the effect of body mass index (BMI) on breast cancer progression \cite{guo2017}. In order to be included in an analysis of disease progression, a participant must have had an initial disease event. If BMI is a risk factor for breast cancer risk, then selection into the sample population would be a function of a collider (namely, BMI), and hence bias would occur. Bias would also occur for analysis of a recurrent disease event if the risk factor was a cause of the first disease event. For example, lipoprotein(a) [Lp(a)] levels are known to be associated with the risk of coronary heart disease (CHD). However, a recent study conducted on individuals with already established CHD reported that Lp(a) was not associated with future cardiovascular mortality \cite{zewinger2017}. We return to this example later in the paper to assess whether this result could be explained by selection bias.

\subsection*{Assessing the causal effect of a risk factor in elderly population}
Another form of selection bias is survivor bias, which may occur when considering a disease outcome in an elderly population \cite{vansteelandt2017}. For example, a recent Mendelian randomization investigation considered the effect of BMI on Parkinson's disease risk \cite{noyce2017}. Genetic associations with disease risk were estimated in a consortium of studies with mean age varying from 48.9 to 76.2 years. While selection due to mortality may be negligible in a study of young adults, it would be more concerning in an elderly sample population. As above, since BMI is a risk factor for increased mortality, selection bias could occur.

\subsection*{Assessing a causal effect in a subpopulation}
Selection bias could also occur when considering causal effects in a subset of the population. For example, genetic associations of alcohol related variants with oesophageal cancer have been considered separately in non-drinkers, moderate drinkers, and heavy drinkers \cite{lewis2005b} -- stronger associations would be expected in heavier drinkers. Selection bias would have occurred here, as strata were defined based on the exposure. In contrast, sex-stratified analyses of alcohol related variants \cite{cho2015} should not be affected by selection bias, as sex is determined at conception and cannot be an effect of any other variable (and so is not a collider).

\section*{Simulation study}
To investigate the magnitude of bias and Type 1 error inflation in a typical Mendelian randomization investigation, we perform a simulation study. We start with a base scenario, and then vary different parameters of the data-generating model in turn.

\subsection*{Data-generating model and choice of parameters}
Our simulation model is as follows (individuals are indexed by $i$):
\begin{align}
X_i  &= \alpha_G \; G_i + \alpha_U \; U_i + \sqrt{1 - \alpha_G^2 - \alpha_U^2} \; \epsilon_{Xi} \notag \\
Y_i  &=  \beta_X \; X_i +  \beta_U \; U_i + \sqrt{1 -  \beta_X^2 -  \beta_U^2} \; \epsilon_{Yi} \notag \\
S_i &\sim \mbox{Bernoulli}(\pi_i) \notag \\
\logit(\pi_i) &= \gamma_0 + \gamma_X X_i + \gamma_U U_i \notag \\
G_i, U_i, \epsilon_{Xi}, \epsilon_{Yi} &\sim \mathcal{N}(0, 1) \mbox{ independently} \notag
\end{align}
The risk factor ($X$) is a linear combination of the genetic variant ($G$), confounder ($U$), and an independent error term ($\epsilon_X$). The outcome ($Y$) is a linear combination of the risk factor, confounder, and an independent error term ($\epsilon_Y$). Selection ($S$) is modelled as a Bernoulli trial with probability of selection $\pi$, which is assumed to be a logistic-linear function of the risk factor and confounder. The parameter $\beta_X$ denotes the causal effect to be estimated.

We have specified the model so that the risk factor and the outcome both have standard normal distributions. Consequently, the parameters $\alpha_G^2$, $\alpha_U^2$ can be interpreted as the proportion of variance in the risk factor that can be explained by the genetic instrument and confounder respectively, while $\beta_X^2$ and $\beta_U^2$ have a similar interpretation for the outcome. Finally, the constant term $\gamma_0$ determines the prevalence of the selection event ($S = 1$).

We consider six different simulation scenarios. In the baseline Scenario 1, the parameters are specified as follows. We set $\alpha_G = \sqrt{0.02}$, $\alpha_U = \sqrt{0.5}$, $\beta_U = \sqrt{0.5}$, meaning that the genetic variant explains 2\% of the variance in the risk factor, and the confounder explains 50\% of the variance in both the risk factor and the outcome. We assume a null causal effect of the risk factor on the outcome ($\beta_X = 0$). We also set $\gamma_0 = 0$ and $\gamma_U = 0$, which corresponds to setting the median probability of selection to $0.5$ and assuming that the confounder does not influence selection. The risk factor effect on selection $\gamma_X$ is allowed to take values $-2, -1, -0.5, -0.2, 0, 0.2, 0.5, 1, 2$. The odds of selection per 1 standard deviation increase in the risk factor is $\exp(\gamma_X)$; for $\gamma_X = 2$, there is a $\exp(2) = 7.39$-fold increase in the odds of selection per standard deviation increase in the risk factor.

We then vary in turn: the proportion of variance in the risk factor explained by the genetic variant $\alpha_G = \sqrt{0.01}, \sqrt{0.05}, \sqrt{0.1}$ (Scenario 2); the proportion of variance in the risk factor explained by the confounder $\alpha_U = \sqrt{0.2}, \sqrt{0.8}$ (Scenario 3); the proportion of variance in the outcome explained by the confounder $\beta_U = \sqrt{0.2}, \sqrt{0.8}$ (Scenario 4); the effect of the confounder on selection $\gamma_U = -1, 1$ (Scenario 5); and the probability of selection $\gamma_0 = -1, -2, -2.4$ (Scenario 6).

We simulate data on a population of 100\thinspace000 individuals, and then randomly select 10\thinspace000 individuals with $S=1$ as the sample. In Scenario 6, for $\gamma_0 = -2$, the analysis is based on a sample size of 1500 instead, as the median probability of selection is 2.3\%. For $\gamma_0 = -2.4$, the analysis is based on a sample size of 500, as the median probability is 0.8\%. 10\thinspace000 simulated datasets are generated for each set of parameters. In each simulation, we estimate the causal effect of the risk factor on the outcome using linear regression for the instrument--risk factor and instrument--outcome associations and computing the ratio estimate, $\hat{\beta}_X = \frac{\hat{\beta}_{Y | G}}{\hat{\beta}_{X | G}}$.

\subsection*{Results}
Results are provided in Table~\ref{scenone} (Scenario 1) and Table~\ref{scenothers} (other scenarios). In Table~\ref{scenone}, we report the mean, median and standard deviation for the estimated effect of the risk factor on the outcome, the median standard error of these effect estimates, and the empirical Type 1 error rate for the risk factor--outcome association at a 5\% level of significance level (defined as the proportion of datasets for which $|\frac{\hat{\beta}_{X}}{\se(\hat{\beta}_{X})}| > 1.96$). In Table~\ref{scenothers}, we consider a slightly narrower range of values for the selection effect, and only provide the median causal effect estimates and empirical Type 1 error rates.

In Scenario 1, when the effect $\gamma_X$ of the risk factor on selection is weak, the mean causal effect estimates are nearly unbiased. However, as the strength of the selection effect $\gamma_X$ increases, so does the magnitude of bias. In the rather extreme case where $\gamma_X = \pm 2$, bias is so large that the null causal hypothesis is rejected in almost $80\%$ of the simulations.

The direction of selection bias in Table~\ref{scenone} is negative, regardless of the direction or magnitude of the risk factor--selection parameter. The direction of selection bias depends on the confounder effects $\alpha_U, \beta_U$ on the risk factor and the outcome. If $\alpha_U$ and $\beta_U$ have the same sign, the causal effect estimate is biased downwards; if not, it is biased upwards (Supplementary Table~\ref{directions1}).

\begin{table}[htbp]
\begin{minipage}{\textwidth}
\begin{center}
\begin{small}
\centering
\begin{tabular}[c]{cc|ccccc}
\hline
$\gamma_X$  & Odds ratio  &  Mean   & Median  	& SD & Med SE  	&  Type 1 error rate (\%) \\
\hline
$-2$        & 0.14        & -0.296 	& -0.289	& 0.123  & 0.106	& 77.7 \% 	\\
$-1$        & 0.37        & -0.107 	& -0.103 	& 0.089  & 0.083	& 24.3 \% 	\\
$-0.5$      & 0.61        & -0.032 	& -0.029 	& 0.077  & 0.074	&  6.6 \% 	\\
$-0.2$      & 0.82        & -0.007 	& -0.004 	& 0.072  & 0.071	&  5.0 \% 	\\
0           & 1.00        & -0.002 	&  0.000 	& 0.071  & 0.071	&  5.1 \% 	\\
0.2         & 1.22        & -0.007 	& -0.004 	& 0.072  & 0.071	&  4.8 \% 	\\
0.5         & 1.65        & -0.032 	& -0.030 	& 0.076  & 0.074	&  6.6 \% 	\\
1           & 2.72        & -0.107 	& -0.103 	& 0.089  & 0.083  	& 23.6 \% 	\\
2           & 7.39        & -0.296 	& -0.289 	& 0.123  & 0.106	& 77.9 \% 	\\
\hline
\end{tabular}
\caption{Mean, median, standard deviation (SD), median standard error (Med SE) of estimates and empirical Type 1 error rate (\%) at a 5\% level of significance for associations of the risk factor with the outcome in Scenario 1, for different values of the selection effect ($\gamma_X$, also expressed as the odds ratio per 1 standard deviation increase in the risk factor).} \label{scenone}
\end{small} %
\end{center}
\end{minipage}
\end{table}
\setlength{\tabcolsep}{6pt}

In the five scenarios of Table~\ref{scenothers}, we investigated the impact of varying different parameters on the magnitude of selection bias. In Scenario 2, we varied instrument strength. This did not affect the magnitude of selection bias (see also \cite{hughes2017}). However, stronger instruments led to larger standard errors and hence increased Type 1 error rates. Weak instrument bias is unlikely to have affected our simulations, since we used a single genetic instrument and weak instrument bias is usually small in this case \cite{burgess2010avoiding}. Also, even with $\alpha_G = \sqrt{0.01}$, the average F statistic for regression of the risk factor on the instrument was around 100.

In Scenarios 3 and 4, we varied the parameters $\alpha_U$ and $\beta_U$, representing the confounder effects on the risk factor and the outcome respectively. In both cases, we observed a moderate increase in the magnitude of selection bias as the strength of the confounder effect increased.

In Scenario 5, we considered a selection procedure influenced by both the risk factor and the confounder. Selection bias is present in this scenario, but the direction of bias also depends on the confounder--selection parameter $\gamma_U$ in a non-linear and non-monotonic way. In the simulations of Table~\ref{scenothers}, the causal effect is underestimated if the confounder and risk factor effects on selection have the same direction. It is mildly overestimated if the risk factor and confounder affect the probability of selection in opposite directions, except when the effect of the risk factor is significantly stronger than that of the confounder, in which case the causal effect is again underestimated. These results also depend on the directions of effects of the confounder on the risk factor and the outcome (Supplementary Table~\ref{directions2}).

Finally in Scenario 6, there was a weak effect of the probability of selection on selection bias. In particular, bias was slightly reduced when selection was less common. Type 1 error rates also reduced since simulations for less common selection were based on a smaller sample size, resulting in larger standard errors.

\begin{table}[htbp]
\begin{minipage}{\textwidth}
\begin{center}
\begin{small}
\centering
\begin{tabular}[c]{c|cc|cc|cc}
\hline
$\gamma_X$  &  Median   & Type 1 error   &  Median   & Type 1  error  &  Median   & Type 1  error  \\
\hline
Scenario 2: & \multicolumn{2}{c|}{$\alpha_G = \sqrt{0.01}$}
                                       & \multicolumn{2}{c|}{$\alpha_G = \sqrt{0.05}$}
                                                                  & \multicolumn{2}{c}{$\alpha_G = \sqrt{0.1}$}  \\
\hline
$-1$        & -0.101 & 13.9 \% 	& -0.104 & 50.4 \% 	& -0.103 & 79.3 \% 	\\
$-0.5$      & -0.030 &  5.9 \% 	& -0.030 &  9.8 \% 	& -0.029 & 14.1 \% 	\\
$-0.2$      & -0.004 &  5.2 \%	& -0.005 &  5.0 \% 	& -0.005 &  5.3 \% 	\\
0           & -0.001 &  5.0 \%	& -0.001 &  5.1 \% 	&  0.000 &  4.9 \% 	\\
0.2         & -0.006 &  5.3 \%	& -0.005 &  5.2 \% 	& -0.005 &  5.4 \% 	\\
0.5         & -0.027 &  5.6 \%	& -0.029 &  9.8 \% 	& -0.029 & 13.8 \% 	\\
1           & -0.104 & 14.0 \%	& -0.103 & 49.9 \% 	& -0.102 & 79.7 \% 	\\
\hline
Scenario 3: & \multicolumn{2}{c|}{$\alpha_U = \sqrt{0.2}$}
                                       & \multicolumn{2}{c|}{$\alpha_U = \sqrt{0.5}$}
                                                                  & \multicolumn{2}{c}{$\alpha_U = \sqrt{0.8}$}  \\
\hline
$-1$        & -0.064 & 12.1 \%	& -0.105 & 24.3 \%	& -0.130 & 35.1 \%	\\
$-0.5$      & -0.018 &  5.7 \%	& -0.030 &  6.6 \%	& -0.039 &  8.0 \%	\\
$-0.2$      & -0.003 &  4.6 \%	& -0.005 &  5.4 \%	& -0.006 &  5.1 \%	\\
0           &  0.002 &  4.9 \%	&  0.000 &  4.8 \%	&  0.000 &  5.2 \%	\\
0.2         & -0.004 &  4.8 \%	& -0.005 &  5.4 \%	& -0.007 &  5.1 \%	\\
0.5         & -0.021 &  5.6 \%	& -0.029 &  6.6 \%	& -0.038 &  7.9 \%	\\
1           & -0.067 & 12.2 \%	& -0.103 & 24.4 \%	& -0.131 & 35.8 \%	\\
\hline
Scenario 4: & \multicolumn{2}{c|}{$\beta_U  = \sqrt{0.2}$}
                                       & \multicolumn{2}{c|}{$\beta_U  = \sqrt{0.5}$}
                                                                  & \multicolumn{2}{c}{$\beta_U  = \sqrt{0.8}$}  \\
\hline
$-1$        & -0.065 & 11.8 \% 	& -0.104 & 24.2 \% 	& -0.131 & 35.5 \% 	\\
$-0.5$      & -0.019 &  5.7 \% 	& -0.029 &  6.4 \% 	& -0.038 &  7.9 \% 	\\
$-0.2$      & -0.002 &  5.0 \% 	& -0.005 &  5.1 \% 	& -0.007 &  4.6 \% 	\\
0           &  0.000 &  5.3 \% 	& -0.001 &  4.9 \% 	&  0.000 &  4.9 \% 	\\
0.2         & -0.002 &  5.1 \% 	& -0.003 &  4.9 \% 	& -0.005 &  5.2 \% 	\\
0.5         & -0.018 &  5.4 \% 	& -0.029 &  6.6 \% 	& -0.039 &  8.0 \% 	\\
1           & -0.065 & 12.1 \% 	& -0.100 & 22.7 \% 	& -0.129 & 34.8 \% 	\\
\hline
Scenario 5: & \multicolumn{2}{c|}{$\gamma_U = -1$}
                                       & \multicolumn{2}{c|}{$\gamma_U = 0$}
                                                                  & \multicolumn{2}{c}{$\gamma_U = 1$}           \\
\hline
$-2$		& -0.293 & 87.4 \%	& -0.290 & 78.3 \%  & -0.110 & 18.1 \%  \\
$-1$        & -0.145 & 45.3 \%	& -0.103 & 24.0 \%	&  0.043 &  8.9 \%	\\
$-0.5$      & -0.069 & 16.0 \%	& -0.028 &  6.9 \%	&  0.043 & 10.0 \%	\\
$-0.2$      & -0.025 &  6.6 \%	& -0.004 &  5.4 \%	&  0.023 &  6.3 \%	\\
0           &  0.002 &  4.9 \%	&  0.000 &  5.0 \%	& -0.001 &  5.5 \%	\\
0.2         &  0.023 &  6.4 \%	& -0.005 &  4.8 \%	& -0.025 &  6.3 \%	\\
0.5         &  0.046 &  9.7 \%	& -0.029 &  6.4 \%	& -0.068 & 15.0 \%	\\
1           &  0.042 &  9.1 \%	& -0.101 & 23.2 \%	& -0.146 & 45.3 \%	\\
2			& -0.112 & 18.6 \%  & -0.291 & 77.7 \%  & -0.293 & 87.1 \%  \\
\hline
Scenario 6: & \multicolumn{2}{c|}{$\gamma_0 = -1$}
                                       & \multicolumn{2}{c|}{$\gamma_0 = -2$}
                                                                  & \multicolumn{2}{c}{$\gamma_0 = -2.4$}        \\
\hline
$-1$        & -0.103 & 23.5 \%	& -0.086 &  6.7 \%	& -0.064 &  5.4 \%	\\
$-0.5$      & -0.024 &  6.4 \%	& -0.019 &  4.8 \%	&  0.000 &  5.0 \%	\\
$-0.2$      & -0.007 &  4.9 \%	& -0.002 &  5.0 \%	& -0.001 &  4.9 \%	\\
0           &  0.001 &  4.4 \%	& -0.002 &  5.2 \%	& -0.006 &  4.9 \%	\\
0.2         & -0.003 &  5.2 \%	&  0.000 &  4.9 \%	& -0.002 &  5.0 \%	\\
0.5         & -0.027 &  6.3 \%	& -0.018 &  4.9 \%	& -0.012 &  5.4 \%	\\
1           & -0.104 & 24.1 \%	& -0.081 &  6.9 \%	& -0.072 &  5.7 \%	\\
\hline
\end{tabular}
\caption{Median association of the risk factor with the outcome and empirical Type 1 error rate (\%) in Scenario 2 (varying instrument strength), Scenario 3 (varying confounder effect on risk factor), Scenario 4 (varying confounder effect on outcome), Scenario 5 (varying confounder effect on selection probability), and Scenario 6 (varying prevalence of selection) for different values of the selection effect ($\gamma_X$).} \label{scenothers}
\end{small}
\end{center}
\end{minipage}
\end{table}
\setlength{\tabcolsep}{6pt}

The scenarios that we have considered are by no means exhaustive. Additional scenarios are reported in the Supplementary Material. When selection depends only on the risk factor, we observed that the magnitude of selection bias is independent of the true value of the risk factor--outcome causal effect (Supplementary Table~\ref{scenseven}). When selection is influenced by the outcome only, or by the outcome and confounder, estimates are unbiased when the true causal effect is null, and only biased when there is a causal effect of the risk factor on the outcome (under the causal null, selection is not downstream of the genetic variant and so not a collider, Supplementary Table~\ref{outcome}). Finally, selection bias acted similarly in simulations with a binary outcome as with a continuous outcome (Supplementary Table~\ref{scenbinary}).

In each of the scenarios presented, bias and Type 1 error rate inflation were minimal when $\gamma_X = \pm 0.2$ (that is, each additional standard deviation increase/decrease in the risk factor led to around a 20\% greater/lower chance of selection). Bias and Type 1 error rate inflation were minimal with $\gamma_X = \pm 0.5$ (65\% greater/40\% lower chance per standard deviation increase/decrease in risk factor) in all scenarios except Scenario 2 with $\alpha_G = \sqrt{0.1}$ and Scenario 5, in which the confounder also affected selection. While these simulation findings are not universally applicable, in particular the extent of Type 1 error inflation (which would be more severe if the sample size was much bigger, or the instrument was much stronger), they suggest that moderate selection bias is unlikely to have a serious impact on moderate-sized Mendelian randomization investigations. In comparison, moderate levels of pleiotropy have been shown to lead to more severe bias and Type 1 error inflation \cite{bowden2015, bowden2015median}.

\section*{Inverse probability weighting}
One common solution to the problem of selection bias is to inversely weight the sample population by the probability of selection into the sample \cite{seaman2011, hernan2006b}. The intuition is that individuals with low probability of selection are likely to be underrepresented in the sample. Inverse probability weighting upweights these individuals to account for other individuals with similar characteristics in the population that were not included in the sample. For example, if an individual included in the sample population would have been sampled with 100\% probability, then that individual does not need to be upweighted, whereas if a selected individual would have been sampled with 20\% probability, that individual is effectively replicated four times to represent the 80\% of similar individuals who were not sampled.

\subsection*{Simulations with inverse probability weighting}
To investigate the utility of inverse probability weighting to correct for selection bias in Mendelian randomization, we extend the simulations presented in the previous section. We consider Scenario 5 with a varying confounder effect on selection, $\gamma_U = -1, 0, 1$, where $\gamma_U = 0$ is equivalent to Scenario 1. We perform logistic regression of selection on the risk factor in the full population of 100\thinspace000 individuals to estimate the selection probabilities, and then perform linear regression of the outcome on the genetic variant weighting by the reciprocals of these probabilities in the 10\thinspace000 selected individuals only. For $\gamma_U = 0$, the selection model is correctly specified, while for $\gamma_U = \pm 1$ it is not.

\subsection*{Trimming of weights}
A disadvantage of inverse probability weighting is that individuals with a very small probability of selection can receive a very large weight in the analysis. While this is appropriate theoretically, the presence of such individuals can lead to highly variable and imprecise estimates. It is common in practice to trim weights above some threshold \cite{lee2011} -- for example, to set the largest $1\%$ of weights to be equal to the 99th percentile of the empirical distribution of weights. In our simulations, we perform analyses with no trimming, and with trimming at the 99th and 95th percentiles.

\subsection*{Results}
Simulations were repeated for 10\thinspace000 datasets for each set of parameters. The results are shown in Table~\ref{scentrim}. When the inverse probability model was correctly specified ($\gamma_U = 0$), inverse probability weighting reduced selection bias, and the resulting causal effect estimates were unbiased. When the weighting model was not correctly specified ($\gamma_U = \pm1$), bias was present. For small values of $\gamma_X$, bias induced by inverse probability weighting was worse than that arising from selection bias. For large values of $\gamma_X$, inverse probability weighting did improve bias, even though the weighting model was not correctly specified. In practice, additional information on possible confounders is often available and can also be incorporated in the weighting scheme. Somewhat paradoxically, although increasing the effect of the confounder on the risk factor $\alpha_U$ increases selection bias, it also increases the correlation between the risk factor and confounder, meaning that misspecification in the weighted model based on the risk factor only is less severe (Supplementary Table~\ref{sceniptw}). Trimming had little effect on results except in the case of extreme values of the risk factor--selection parameter $\gamma_X = \pm2$, where it re-introduced some of the bias that had been removed by using inverse probability weighting, but reduced the variability of estimates.

\setlength{\tabcolsep}{4pt}
\begin{table}[htbp]
\begin{minipage}{\textwidth}
\begin{adjustwidth}{-0.4cm}{-0.4cm}
\begin{center}
\begin{footnotesize}
\centering
\begin{tabular}[c]{c|cccc|cccc|cccc}
\hline
$\gamma_X$  &  Median   &   SD    &  Med SE  &  Type 1 &  Median   &   SD    &  Med SE  &  Type 1 &  Median   &   SD    &  Med SE  &  Type 1   \\
\hline
$\gamma_U = 0$
            & \multicolumn{4}{c|}{No trimming}        & \multicolumn{4}{c|}{Trimming at 99\%}   & \multicolumn{4}{c}{Trimming at 95\%}      \\
\hline
$-2$        & -0.008 & 6.499 & 0.072 & 39.6 \% & -0.113 & 0.129 & 0.085 & 33.8 \% & -0.206 & 0.124 & 0.096 & 56.8 \% \\
$-1$        & -0.002 & 0.091 & 0.071 & 11.4 \% & -0.032 & 0.089 & 0.075 & 10.7 \% & -0.076 & 0.091 & 0.080 & 17.8 \% \\
$-0.5$      & -0.002 & 0.076 & 0.071 &  6.3 \% & -0.010 & 0.076 & 0.072 &  6.3 \% & -0.027 & 0.078 & 0.074 &  7.2 \% \\
$-0.2$      &  0.000 & 0.072 & 0.071 &  5.2 \% & -0.002 & 0.072 & 0.071 &  5.1 \% & -0.007 & 0.073 & 0.072 &  5.1 \% \\
0           &  0.001 & 0.072 & 0.071 &  5.0 \% &  0.001 & 0.072 & 0.071 &  5.0 \% &  0.001 & 0.072 & 0.071 &  5.0 \% \\
0.2         &  0.001 & 0.072 & 0.071 &  5.0 \% & -0.001 & 0.072 & 0.071 &  4.9 \% & -0.006 & 0.073 & 0.072 &  5.1 \% \\
0.5         &  0.001 & 0.076 & 0.071 &  6.5 \% & -0.008 & 0.076 & 0.072 &  6.4 \% & -0.024 & 0.078 & 0.074 &  6.7 \% \\
1           & -0.001 & 0.091 & 0.071 & 11.3 \% & -0.032 & 0.089 & 0.075 & 10.7 \% & -0.074 & 0.092 & 0.080 & 17.8 \% \\
2           & -0.008 & 0.902 & 0.072 & 38.8 \% & -0.118 & 0.130 & 0.085 & 34.2 \% & -0.210 & 0.125 & 0.096 & 58.1 \% \\
\hline
$\gamma_U = -1$
            & \multicolumn{4}{c|}{No trimming}        & \multicolumn{4}{c|}{Trimming at 99\%}   & \multicolumn{4}{c}{Trimming at 95\%}      \\
\hline
$-2$        & -0.031 & 1.226 & 0.058 & 49.0 \% & -0.130 & 0.109 & 0.071 & 47.3 \% & -0.207 & 0.103 & 0.081 & 69.5 \% \\
$-1$        &  0.009 & 0.110 & 0.058 & 24.0 \% & -0.043 & 0.086 & 0.065 & 17.6 \% & -0.097 & 0.086 & 0.072 & 30.7 \% \\
$-0.5$      &  0.025 & 0.076 & 0.059 & 14.7 \% & -0.003 & 0.075 & 0.063 &  9.5 \% & -0.040 & 0.077 & 0.068 & 11.6 \% \\
$-0.2$      &  0.033 & 0.069 & 0.061 & 11.9 \% &  0.016 & 0.069 & 0.063 &  7.9 \% & -0.010 & 0.072 & 0.067 &  6.7 \% \\
0           &  0.040 & 0.067 & 0.063 & 11.8 \% &  0.029 & 0.067 & 0.064 &  8.8 \% &  0.010 & 0.069 & 0.066 &  6.0 \% \\
0.2         &  0.043 & 0.066 & 0.064 & 10.6 \% &  0.037 & 0.066 & 0.065 &  9.1 \% &  0.024 & 0.068 & 0.067 &  6.8 \% \\
0.5         &  0.049 & 0.067 & 0.067 & 10.9 \% &  0.047 & 0.067 & 0.068 & 10.5 \% &  0.043 & 0.068 & 0.068 &  9.7 \% \\
1           &  0.050 & 0.074 & 0.074 & 10.9 \% &  0.047 & 0.074 & 0.074 & 10.3 \% &  0.041 & 0.075 & 0.075 &  8.9 \% \\
2           &  0.032 & 0.123 & 0.086 & 16.9 \% & -0.013 & 0.117 & 0.093 & 11.0 \% & -0.067 & 0.119 & 0.100 & 13.9 \% \\
\hline
$\gamma_U = 1$
            & \multicolumn{4}{c|}{No trimming}        & \multicolumn{4}{c|}{Trimming at 99\%}   & \multicolumn{4}{c}{Trimming at 95\%}      \\
\hline
$-2$        &  0.030 & 0.122 & 0.087 & 16.7 \% & -0.015 & 0.117 & 0.093 & 10.6 \% & -0.070 & 0.119 & 0.100 & 13.8 \% \\
$-1$        &  0.052 & 0.072 & 0.073 & 10.9 \% &  0.049 & 0.072 & 0.074 & 10.1 \% &  0.042 & 0.073 & 0.075 &  8.6 \% \\
$-0.5$      &  0.047 & 0.067 & 0.067 & 11.0 \% &  0.045 & 0.068 & 0.067 & 10.5 \% &  0.041 & 0.068 & 0.068 &  9.5 \% \\
$-0.2$      &  0.045 & 0.067 & 0.064 & 11.9 \% &  0.039 & 0.067 & 0.065 & 10.0 \% &  0.026 & 0.069 & 0.067 &  7.4 \% \\
0           &  0.039 & 0.066 & 0.062 & 11.4 \% &  0.028 & 0.067 & 0.064 &  8.6 \% &  0.009 & 0.069 & 0.066 &  6.1 \% \\
0.2         &  0.033 & 0.070 & 0.061 & 12.0 \% &  0.016 & 0.070 & 0.063 &  8.1 \% & -0.011 & 0.072 & 0.067 &  6.9 \% \\
0.5         &  0.025 & 0.076 & 0.060 & 14.1 \% & -0.004 & 0.074 & 0.063 &  9.1 \% & -0.042 & 0.076 & 0.068 & 11.5 \% \\
1           &  0.005 & 0.102 & 0.058 & 24.1 \% & -0.047 & 0.085 & 0.065 & 17.9 \% & -0.100 & 0.086 & 0.072 & 31.0 \% \\
2           & -0.034 & 1.709 & 0.058 & 48.5 \% & -0.132 & 0.110 & 0.071 & 48.0 \% & -0.209 & 0.104 & 0.081 & 70.2 \% \\
\hline
\end{tabular}
\caption{Median, standard deviation (SD), median standard error (Med SE) of estimates and empirical Type 1 error rate (\%) for the risk factor-outcome causal effect with correctly specified inverse probability weighting selection model ($\gamma_U = 0$) and misspecified selection model ($\gamma_U = \pm 1$) for different values of the selection effect ($\gamma_X$).} \label{scentrim}
\end{footnotesize} %
\end{center}
\end{adjustwidth}
\end{minipage}
\end{table}
\setlength{\tabcolsep}{6pt}

\section*{Example: effect of lipoprotein(a) on secondary cardiovascular disease}
\label{sec:lpa}
Lp(a) is an unusual risk factor for Mendelian randomization as genetic variants in the \emph{LPA} gene region explain up to $90\%$ of its variance \cite{boerwinkle1992}. This comes in contrast to most Mendelian randomization investigations, where genetic variants explain a small proportion of the variance in the risk factor, typically about $1-4\%$. Consequently, even moderate selection bias may have a serious impact on Mendelian randomization analyses of Lp(a). Previous investigations have demonstrated associations between genetic variants in the \emph{LPA} gene region and coronary heart disease \cite{clarke2009lpa, kamstrup2009}. However, a recent investigation of individuals with previous established coronary heart disease did not find an association between variants in the same region and subsequent cardiovascular mortality \cite{zewinger2017}. We consider by simulation whether this result could be explained by selection bias.

Our data-generating model is the same as in the simulation study except that the outcome is binary:
\begin{align}
	Y_i &\sim  \text{Bernoulli} (\pi_{Yi}) \nonumber \\
	\logit \pi_{Yi} &= \beta_0 + \beta_X X_i + \beta_U U_i. \nonumber
\end{align}
Parameter values are informed by the literature on Lp(a) to resemble the analysis of Zewinger et al.\ \cite{zewinger2017}, with the selection variable $S$ representing an initial CHD event and the outcome $Y$ representing cardiovascular mortality. As in Zewinger et al.\ \cite{zewinger2017}, we use a sample size of $n = 3313$ for the main analysis. This is assumed to be drawn from a larger population of size $N = 100000$. We use a single genetic instrument which explains 36\% of the variation in Lp(a) levels ($\alpha_G = \sqrt{0.36}$); this is the proportion of variation previously reported \cite{clarke2009lpa} as explained by the two variants associated with Lp(a) levels that were used in Zewinger et al.'s analysis. We also set $\alpha_U = \sqrt{0.32}$ implying that half of the remaining variation in Lp(a) is due to the confounder. We assume that the effect of Lp(a) on CHD risk (the selection event) is equal to the effect of Lp(a) on cardiovascular mortality (the outcome event): $\gamma_X = \beta_X = +0.25$. Similarly the effects of the confounder on CHD risk ($\gamma_U$) and cardiovascular mortality ($\beta_U$) are assumed equal. We set $\gamma_0 = -2$ meaning that around 20\% of the population experience a CHD event and survive to be eligible for selection. We set $\beta_0$ to obtain around 20\% outcome events in the selected sample (corresponding to the 621 cardiovascular deaths in the original study). We took different values of the confounder effects $\gamma_U = \beta_U = 0, +0.2, +0.5, +1, +1.5, +2$. We generated 10\thinspace000 datasets for each value of the confounder effect, and calculated in each case the association coefficient from logistic regression for the first 3313 participants in the population (no selection), and the first 3313 with the selection event.

\begin{table}[htbp]
\begin{minipage}{\textwidth}
\begin{center}
\begin{small}
\centering
\begin{tabular}[c]{cc|c|cc}
\hline
                      &           & No selection   & \multicolumn{2}{c}{With selection}  \\
$\beta_U$, $\gamma_U$ & $\beta_0$ & Mean estimate  &  Mean estimate   & Empirical power  \\
\hline
0                     & $-1.4$    &  0.149         &  0.149           &  93.5\%          \\
+0.2                  & $-1.6$    &  0.148         &  0.145           &  91.3\%          \\
+0.5                  & $-1.9$    &  0.142         &  0.133           &  86.1\%          \\
+1                    & $-2.5$    &  0.131         &  0.102           &  67.7\%          \\
+1.5                  & $-3.3$    &  0.120         &  0.077           &  44.0\%          \\
+2                    & $-4.0$    &  0.107         &  0.061           &  30.4\%          \\
\hline
\end{tabular}
\caption{Mean association estimates in the population (``no selection") and among individuals with a CHD event (``with selection"), and empirical power at a 5\% level of significance for different magnitudes of confounding in the applied example. (The $\beta_0$ parameter is chosen such that the proportion of cases in the selected sample is about 20\% for each value of $\beta_U$ and $\gamma_U$.)} \label{lparesults}
\end{small} %
\end{center}
\end{minipage}
\end{table}
\setlength{\tabcolsep}{6pt}

Table~\ref{lparesults} shows the results: the mean association estimates with no selection and with selection, and the empirical power under selection. Empirical power for the sample with no selection was not comparable as there were fewer events in this sample. Even with no selection, the association estimate differed from $\alpha_G \times \beta_X = 0.15$ and attenuated as $\gamma_U$ increased due to non-collapsibility \cite{greenland1999, burgess2012noncollapse}. Bias in mean association estimates due to selection was towards the null, and increased with $\gamma_U$. Although the investigation was well-powered in the absence of selection bias, when $\gamma_U = +1.5$, bias was fairly severe and the empirical power was 45.5\%. When $\gamma_U = +2$, the mean association estimate had reduced by almost half compared to the estimate with no selection, and empirical power was only 31.0\%. While these values of $\gamma_U$ are fairly large, there are individual cardiovascular risk factors such as LDL-cholesterol that are positively correlated with Lp(a) and do have large effects on CHD risk and cardiovascular mortality; a 30\% lowering (approximately 1 standard deviation) of genetically-predicted LDL-cholesterol has previously been shown to reduce CHD risk by 67\% \cite{burgess2013genepi}, corresponding to $\gamma_U = +1.11$ ($-\log(0.33) = +1.11$). When all confounders are considered together, the value of $\gamma_U$ would be larger still. In conclusion, this simulation exercise suggests that it is plausible that the null finding of Zewinger et al.\ may have been obtained due to selection bias.

\section*{Discussion}
The aim of this paper was to consider selection bias in the context of Mendelian randomization. We discussed scenarios in which selection bias may occur, in particular those which are likely to affect Mendelian randomization investigations. We simulated data to be representative of a typical Mendelian randomization investigation, and showed that selection bias can significantly influence causal effect estimates when selection into the study is strongly influenced by the risk factor. However, moderate selection bias did not adversely affect estimates too severely across a range of realistic scenarios. A similar conclusion was reached previously for genetic association estimates in the context of secondary events \cite{hu2017}. Aside from the risk factor--selection parameter, the magnitude of selection bias was shown to be influenced by the strength of the confounder--risk factor and confounder--outcome effects, as well as the confounder--selection parameter and selection frequency. We demonstrated that inverse probability weighting can ameliorate selection bias, but only in cases where the probability of selection can be modelled accurately. When selection bias was moderate, misspecification in the selection model meant that the `cure could be worse than the disease'. Finally, we considered a somewhat atypical example of a Mendelian randomization analysis in which genetic variants explained a large proportion of variance in the risk factor, and showed that strong (but credibly so) selection bias could explain the anomalous finding that \emph{LPA} variants were not associated with cardiovascular mortality.

Although inverse probability weighting may be helpful in some cases to reduce selection bias, its implementation requires estimation of probability of selection into the study. This typically requires information on individuals who were not included in the study, which may not be available. An important question when considering whether to use inverse probability weighting is to whom the causal estimate relates. As an example, consider estimating the effect of lipid fractions (in particular, LDL-cholesterol) on cognitive performance after a stroke event. A Mendelian randomization analysis of a representative sample of the general population would provide an estimate of the average causal effect of LDL-cholesterol on cognitive performance in the population as a whole (this may be an average treatment effect or a local average treatment effect depending on the precise assumptions made \cite{hernan2006} -- although previous work suggests a Mendelian randomization estimate represents the effect of life-long intervention in a risk factor, and therefore may be a poor guide as to the impact of intervention on the risk factor in a practical setting \cite{burgess2012bmj}). Restricting the estimate to those who had a stroke event is likely to lead to selection bias. By inverse probability weighting, we can potentially resolve the problem of selection bias, but now our estimate is re-weighted back to the original population -- it represents the average effect of intervening on LDL-cholesterol in the population as if everyone in the population had a stroke event. Therefore by inverse probability weighting, we have resolved the problem that the instrumental variable assumptions were violated in the sample population, but now our causal estimate relates to the general population and not the sample population.

In the majority of simulations in this manuscript, we have modelled selection as depending on the risk factor and a single confounder with linear relationships between variables, and the probability of selection as a logistic variable. Although we suspect that our findings will apply to different selection models, it would not be feasible to verify this for every possible model configuration, as well as for binary and time-to-event outcomes. However, our results were robust across a range of realistic scenarios. A potential extension of this work is to develop an analytic bias calculator for instrumental variable analysis. This would be a useful tool for sensitivity analysis not only for Mendelian randomization, but also for other contexts in which instrumental variable analysis is used to analyse observational data.

In conclusion, selection bias can have an adverse effect on Mendelian randomization studies, but in most cases its importance will be less than other sources of bias, such as pleiotropy or population stratification.

\subsection*{Acknowledgements}

This work was supported by the UK Medical Research Council (Core Medical Research Council Biostatistics Unit Funding Code: MC\_UU\_00002/7). Apostolos Gkatzionis was supported by a Medical Research Council Methodology Research Panel grant (Grant Number RG88311). Stephen Burgess was supported by a Sir Henry Dale Fellowship jointly funded by the Wellcome Trust and the Royal Society (Grant Number 204623/Z/16/Z).

\bibliographystyle{vancouver}
\bibliography{masterref}

\newpage
\renewcommand{\thesection}{A\arabic{section}}
\renewcommand{\thesubsection}{A\arabic{subsection}}
\renewcommand{\thesubsubsection}{A.\arabic{subsubsection}}
\renewcommand{\thetable}{A\arabic{table}}
\renewcommand{\thefigure}{A\arabic{figure}}
\renewcommand{\theequation}{A\arabic{equation}}
\renewcommand{\thepage}{A\arabic{page}}
\setcounter{table}{0}
\setcounter{figure}{0}
\setcounter{page}{1}
\setcounter{equation}{0}
\renewcommand{\tablename}{Supplementary Table}
\renewcommand{\figurename}{Supplementary Figure}
\setcounter{section}{0}
\setcounter{subsection}{0}

\section*{Appendix}
\subsection*{Additional simulation examples}
We provide additional simulations to further investigate which aspects of a Mendelian randomization study affect the magnitude of selection bias and the performance of inverse probability weighting.

\subsubsection{Direction of selection bias}
We explored the relationship between the direction of confounder effects on the risk factor and outcome, and the direction of selection bias. For the baseline simulation of Scenario 1, where selection depends only on the risk factor, we varied the signs of these two parameters in our simulations, letting $\alpha_U = \pm \sqrt{0.5}$ and $\beta_U = \pm \sqrt{0.5}$. Results are reported in Supplementary Table~\ref{directions1}. The simulation results indicate that the causal effect is biased downwards if the directions of the confounder effects on the risk factor and the outcome are the same, and upwards otherwise.

\setlength{\tabcolsep}{4pt}
\begin{table}[htbp]
\begin{minipage}{\textwidth}
\begin{adjustwidth}{-0.4cm}{-0.4cm}
\begin{center}
\begin{footnotesize}
\centering
\begin{tabular}[c]{c|cccc|cccc|cccc}
\hline
   & \multicolumn{4}{c|}{$\alpha_U = - \sqrt{0.5}$, $\beta_U = \sqrt{0.5}$}   & \multicolumn{4}{c|}{$\alpha_U = \sqrt{0.5}$, $\beta_U = - \sqrt{0.5}$}   & \multicolumn{4}{c}{$\alpha_U = - \sqrt{0.5}$, $\beta_U = - \sqrt{0.5}$}      \\
$\gamma_X$  &  Median   &   SD    &  Med SE  &  Type 1 &  Median   &   SD    &  Med SE  &  Type 1 &  Median   &   SD    &  Med SE  &  Type 1   \\
\hline
$-2$        &  0.290 & 0.122 & 0.106 & 78.1 \%	&  0.292 & 0.119 & 0.106 & 78.3 \%	& -0.289 & 0.121 & 0.106 & 77.7 \% \\
$-1$        &  0.103 & 0.089 & 0.083 & 23.4 \%	&  0.102 & 0.089 & 0.083 & 23.0 \%	& -0.104 & 0.089 & 0.083 & 24.1 \% \\
$-0.5$      &  0.029 & 0.076 & 0.074 &  7.0 \%	&  0.031 & 0.076 & 0.074 &  6.6 \%	& -0.029 & 0.077 & 0.074 &  7.1 \% \\
$-0.2$      &  0.004 & 0.071 & 0.071 &  4.6 \%	&  0.005 & 0.072 & 0.071 &  5.2 \%	& -0.005 & 0.072 & 0.071 &  5.1 \% \\
0           &  0.000 & 0.070 & 0.071 &  4.8 \%	& -0.001 & 0.072 & 0.071 &  5.1 \%	& -0.001 & 0.071 & 0.071 &  5.0 \% \\
0.2         &  0.006 & 0.072 & 0.071 &  5.0 \%	&  0.005 & 0.073 & 0.071 &  5.3 \%	& -0.005 & 0.072 & 0.071 &  5.1 \% \\
0.5         &  0.029 & 0.077 & 0.074 &  6.7 \%	&  0.029 & 0.077 & 0.074 &  6.9 \%	& -0.028 & 0.075 & 0.074 &  6.6 \% \\
1           &  0.102 & 0.089 & 0.083 & 23.4 \%	&  0.103 & 0.087 & 0.083 & 23.0 \%	& -0.102 & 0.089 & 0.083 & 23.2 \% \\
2           &  0.292 & 0.120 & 0.106 & 78.7 \%	&  0.288 & 0.122 & 0.106 & 77.4 \%	& -0.289 & 0.121 & 0.106 & 77.9 \% \\
\hline
\end{tabular}
\caption{Median, standard deviation (SD), median standard error and 5\% empirical Type 1 error rate for causal effect estimates, for varying directions of the confounder-exposure ($\alpha_U$) and the confounder-outcome ($\beta_U$) effects.} \label{directions1}
\end{footnotesize} %
\end{center}
\end{adjustwidth}
\end{minipage}
\end{table}
\setlength{\tabcolsep}{6pt}

Note that we have made the simplifying assumption that the confounder $U$ represents the cumulative effect of all possible sources of confounding for the risk factor--outcome association, so $\alpha_U$ and $\beta_U$ represent the total effect of all confounders on the risk factor and the outcome. In practice, the signs of these parameters may be difficult to determine if different confounders have opposite effects on the risk factor or the outcome.

We also performed additional simulations, summarized in Supplementary Table~\ref{directions2}, to assess the direction of selection bias when selection depends on both the risk factor and the confounder. For simplicity, we focus only on the direction of bias and ignore its magnitude.

\begin{table}[htbp]
\begin{minipage}{\textwidth}
\begin{center}
\centering
\begin{tabular}[c]{cc|cccc}
\hline
 &  & \multicolumn{2}{c}{$\alpha_U > 0$} & \multicolumn{2}{c}{$\alpha_U < 0$} \\
 &  & $\beta_U > 0$ & $\beta_U < 0$ & $\beta_U > 0$ & $\beta_U < 0$ \\
 \hline
 \multirow{2}{*}{$\gamma_U > 0$} & $\gamma_X > 0$ & $-$ & $+$ & $\mp$ & $\pm$ \\
  & $\gamma_X < 0$ & $\pm$ & $\mp$ & $+$ & $-$ \\
 \multirow{2}{*}{$\gamma_U < 0$} & $\gamma_X > 0$ & $\pm$ & $\mp$ & $+$ & $-$ \\
   & $\gamma_X < 0$ & $-$ & $+$ & $\mp$ & $\pm$ \\
\hline
\end{tabular}
\caption{Direction of selection bias of causal effect estimates when selection depends on the risk factor and the confounder. ``$+$": upward bias, ``$-$": downward bias, ``$\pm$": upward bias for moderate X-S associations, downward bias for strong associations, ``$\mp$": downward bias for moderate X-S associations, upward bias for strong associations.} \label{directions2}
\end{center}
\end{minipage}
\end{table}

A change in the direction of bias (a ``$\pm$" or a ``$\mp$" sign) is observed when the signs of $\gamma_U$ and $\gamma_X \alpha_U$ are different. Intuitively, these parameters express the direct ($\gamma_U$) and indirect ($\gamma_X \alpha_U$, mediated by the risk factor) effect of the confounder on selection. If these two effects act on the same direction, the direction of selection bias is determined again by the effects $\alpha_U$, $\beta_U$ of the confounder on the risk factor and the outcome, as in Supplementary Table~\ref{directions1}. When $\gamma_U$ and $\gamma_X \alpha_U$ have opposite signs, the confounder affects selection in two opposite ways. Selection bias due to the confounder effect as mediated by the risk factor acts in the direction dictated by the $\alpha_U$, $\beta_U$ coefficients, as discussed previously, while selection bias due to the direct effect of the confounder on selection acts in the opposite direction. The relative magnitudes of $\gamma_X \alpha_U$ and $\gamma_U$ determine which effect is stronger, and hence the direction of bias. In the simulations in the main body of the paper (Scenario 5), $\gamma_X$ was the only parameter whose value we varied, so the direction of bias depended on that parameter.

\subsubsection{Selection bias for a non-null causal effect}
To investigate whether selection bias depends on the true value of the risk factor-outcome causal effect, we reproduced the simulations of Table 1 with the causal effect parameter set to $\beta_X = 0.5$ instead of $\beta_X = 0$. Supplementary Table~\ref{scenseven} contains the results of this simulation. The magnitude of selection bias was very similar to that reported in Table~\ref{scenone}. This implies that when selection only depends on the risk factor, the magnitude of selection bias is independent of the value of the causal effect $\beta_X$. Similar results (not reported here) were obtained for a range of different $\beta_X$ values, as well as for a negative causal effect ($\beta_X = - 0.5$).

\begin{table}[htbp]
\begin{minipage}{\textwidth}
\begin{center}
\begin{small}
\centering
\begin{tabular}[c]{cc|ccccc}
\hline
$\gamma_X$  & Odds ratio  &  Mean   & Median  & SD &  Med SE  &  Empirical Power \\
\hline
$-2$        & 0.14        & 0.203 & 0.211 & 0.108  & 0.118 &   42.6 \% \\
$-1$        & 0.37        & 0.392 & 0.396 & 0.078  & 0.098 &   98.1 \% \\
$-0.5$      & 0.61        & 0.466 & 0.468 & 0.066  & 0.090 &   99.9 \% \\
$-0.2$      & 0.82        & 0.492 & 0.494 & 0.061  & 0.087 &  100.0 \% \\
0           & 1.00        & 0.498 & 0.500 & 0.062  & 0.086 &  100.0 \% \\
0.2         & 1.22        & 0.493 & 0.495 & 0.063  & 0.087 &  100.0 \% \\
0.5         & 1.65        & 0.468 & 0.471 & 0.066  & 0.090 &  100.0 \% \\
1           & 2.72        & 0.392 & 0.397 & 0.078  & 0.098 &   98.0 \% \\
2           & 7.39        & 0.205 & 0.211 & 0.108  & 0.119 &   43.2 \% \\
\hline
\end{tabular}
\caption{Mean, median, standard deviation (SD), median standard error and empirical power to reject the null causal hypothesis at a 5\% significance level for causal effect estimates in Scenario 1, with the true causal effect set to $\beta_X = 0.5$.} \label{scenseven}
\end{small} %
\end{center}
\end{minipage}
\end{table}
\setlength{\tabcolsep}{6pt}

\subsubsection{Outcome-dependent selection mechanism}
The selection mechanism used in the simulations of Tables~\ref{scenone} and \ref{scenothers} depended only on the risk factor, except in Scenario 5 where selection also depended on the confounder. Here, we considered an alternative selection procedure, in which selection depends on the outcome and possibly on the confounder.

To implement these simulations, we modified the data-generating model by letting the probability of selection depend on the outcome and the confounder:
\begin{equation*}
	\logit(\pi_i) = \gamma_0 + \gamma_U U_i + \gamma_Y Y_i.
\end{equation*}
Simulations were performed by varying the strength of the outcome--selection parameter $\gamma_Y$, allowing it to take values $-2, -1, -0.5, -0.2, 0, 0.2, 0.5, 1, 2$, and the confounder--selection parameter $\gamma_U$, allowing it to take values $0$ and $+1$. All other parameters were the same as in Scenario 1.

As illustrated in Supplementary Table~\ref{outcome}, there is no selection bias under the null causal hypothesis ($\beta_X = 0$). Additionally in this case, nominal Type 1 error rates are maintained. Therefore a Mendelian randomization study in which selection depends only on the outcome (and possibly on the confounder) will not lead to false positive results. It is possible that a null finding may be a false negative result due to selection bias, but it is somewhat less likely -- it would only occur if selection bias was of the same magnitude as the causal effect and acted in the opposite direction.

\begin{table}[htbp]
\begin{minipage}{\textwidth}
\begin{center}
\begin{small}
\centering
\begin{tabular}[c]{c|cccc|cccc}
\hline
$\gamma_Y$ & \multicolumn{4}{c|}{$\beta_X = 0$}        & \multicolumn{4}{c}{$\beta_X = 0.5$}   \\
\hline
$\gamma_U = 0$  & Median  & SD &  Med SE &   Type 1 Error & Median  & SD &   Med SE &  Emp Power \\
\hline
$-2$        &  0.000 	& 0.057  & 0.056 	& 5.2 \% 	& 0.336 & 0.063  & 0.076 &  99.2 \% \\
$-1$        &  0.001 	& 0.066  & 0.064 	& 5.5 \% 	& 0.420 & 0.062  & 0.082 & 100.0 \% \\
$-0.5$      &  0.001 	& 0.070  & 0.069  	& 5.1 \% 	& 0.474 & 0.061  & 0.085 & 100.0 \% \\
$-0.2$      &  0.000 	& 0.071  & 0.070 	& 5.1 \% 	& 0.495 & 0.061  & 0.086 & 100.0 \% \\
$0$         & -0.001 	& 0.070  & 0.071 	& 4.5 \% 	& 0.500 & 0.062  & 0.086 & 100.0 \% \\
$0.2$       &  0.000 	& 0.071  & 0.070 	& 5.1 \% 	& 0.496 & 0.062  & 0.086 & 100.0 \% \\
$0.5$       &  0.000 	& 0.069  & 0.069 	& 5.2 \% 	& 0.474 & 0.063  & 0.085 & 100.0 \% \\
$1$         &  0.000 	& 0.065  & 0.064 	& 5.3 \% 	& 0.419 & 0.063  & 0.082 &  99.9 \% \\
$2$         &  0.001 	& 0.057  & 0.056 	& 5.2 \% 	& 0.335 & 0.061  & 0.076 &  99.3 \% \\
\hline
$\gamma_U = 1$  & Median  & SD &   Med SE &  Type 1 Error & Median  & SD &    Med SE &  Emp Power  \\
\hline
$-2$        & -0.001 & 0.064  & 0.063 & 5.2 \% 	& 0.328 & 0.069  & 0.086 &  96.8 \% \\
$-1$        & -0.001 & 0.071  & 0.070 & 4.9 \% 	& 0.468 & 0.064  & 0.088 &  99.9 \% \\
$-0.5$      &  0.000 & 0.071  & 0.070 & 5.2 \% 	& 0.512 & 0.060  & 0.085 & 100.0 \% \\
$-0.2$      &  0.000 & 0.069  & 0.069 & 4.9 \% 	& 0.509 & 0.059  & 0.082 & 100.0 \% \\
$0$         &  0.000 & 0.067  & 0.068 & 4.4 \% 	& 0.500 & 0.058  & 0.081 & 100.0 \% \\
$0.2$       &  0.000 & 0.067  & 0.066 & 4.9 \% 	& 0.486 & 0.059  & 0.079 & 100.0 \% \\
$0.5$       & -0.001 & 0.064  & 0.064 & 4.9 \% 	& 0.462 & 0.057  & 0.077 & 100.0 \% \\
$1$         &  0.000 & 0.060  & 0.059 & 4.8 \% 	& 0.423 & 0.057  & 0.074 & 100.0 \% \\
$2$         & -0.001 & 0.055  & 0.053 & 5.6 \% 	& 0.364 & 0.056  & 0.070 & 100.0 \% \\
\hline
\end{tabular}
\caption{Median, standard deviation (SD), median standard error and empirical power to reject the null causal hypothesis at a 5\% significance level (for $\beta_X = 0$, this is equal to the empirical Type 1 error rate) for causal effect estimates where selection depends only on the outcome ($\gamma_U = 0$) or on the outcome and the confounder ($\gamma_U = 1$).} \label{outcome}
\end{small}
\end{center}
\end{minipage}
\end{table}
\setlength{\tabcolsep}{6pt}

\subsubsection{Binary outcomes}

Studying a binary outcome (such as disease status) is quite common in Mendelian randomization analyses. We performed a set of simulations with a binary outcome, using a logistic-linear model to simulate the outcome as in the lipoprotein(a) application. In this case, the causal effect represents the log odds ratio for the outcome per unit increase in the risk factor. In our simulations, we set the causal effect equal to $\beta_X = 0$ and let the remaining parameters take the same values as in Scenario 1. We then varied the constant term $\beta_0$, which dictates the prevalence of the disease outcome in the population, in order to assess whether it has an effect on selection bias.

Results are reported in Supplementary Table~\ref{scenbinary}. We can see that the outcome frequency has practically no effect on selection bias, but a rare event results in increased standard errors for the causal effect estimate. This is not unexpected; for example, the power of a case-control study is maximized when the number of cases and controls is approximately the same. Finally, additional simulations (not reported here) suggested that the performance of inverse probability weighting with a binary outcome is similar to that with a continuous outcome.

\setlength{\tabcolsep}{4pt}
\begin{table}[htbp]
\begin{minipage}{\textwidth}
\begin{adjustwidth}{-0.4cm}{-0.4cm}
\begin{center}
\begin{footnotesize}
\centering
\begin{tabular}[c]{c|cccc|cccc|cccc}
\hline
   & \multicolumn{4}{c|}{$\beta_0 = 0$}   & \multicolumn{4}{c|}{$\beta_0 = - 1.4$}   & \multicolumn{4}{c}{$\beta_0 = - 3$}      \\
$\gamma_X$  &  Median   &   SD    &  Med SE  &  Type 1 &  Median   &   SD    &  Med SE  &  Type 1 &  Median   &   SD    &  Med SE  &  Type 1   \\
\hline
$-2$        & -0.269 & 0.233 & 0.225 & 22.1 \% & -0.279 & 0.305 & 0.295 & 15.9 \% & -0.301 & 0.570 & 0.553 &  8.7 \% \\
$-1$        & -0.093 & 0.173 & 0.171 &  8.5 \% & -0.102 & 0.223 & 0.219 &  7.7 \% & -0.106 & 0.408 & 0.402 &  5.9 \% \\
$-0.5$      & -0.027 & 0.151 & 0.150 &  4.9 \% & -0.030 & 0.189 & 0.187 &  5.4 \% & -0.027 & 0.341 & 0.339 &  5.0 \% \\
$-0.2$      & -0.006 & 0.144 & 0.143 &  5.2 \% & -0.009 & 0.177 & 0.175 &  5.2 \% & -0.008 & 0.318 & 0.313 &  5.2 \% \\
0           & -0.002 & 0.143 & 0.141 &  5.2 \% &  0.000 & 0.172 & 0.171 &  4.9 \% &  0.001 & 0.301 & 0.302 &  4.9 \% \\
0.2         & -0.006 & 0.145 & 0.143 &  5.2 \% &  0.000 & 0.174 & 0.170 &  5.3 \% & -0.001 & 0.304 & 0.299 &  5.2 \% \\
0.5         & -0.027 & 0.153 & 0.150 &  5.6 \% & -0.024 & 0.178 & 0.176 &  4.9 \% & -0.021 & 0.307 & 0.305 &  5.0 \% \\
1           & -0.095 & 0.174 & 0.171 &  8.2 \% & -0.093 & 0.199 & 0.196 &  7.8 \% & -0.100 & 0.343 & 0.336 &  6.4 \% \\
2           & -0.273 & 0.235 & 0.225 & 22.4 \% & -0.260 & 0.256 & 0.251 & 17.4 \% & -0.277 & 0.431 & 0.424 &  9.8 \% \\
\hline
\end{tabular}
\caption{Median, standard deviation (SD), median standard error and 5\% empirical Type 1 error rate for risk factor-outcome causal effect estimates, in simulations with a binary outcome and a varying outcome frequency ($50\%$, $20\%$ and $5\%$, for $\beta_0 = 0, -1.4, -3$ respectively) for different values of the selection effect ($\gamma_X$).} \label{scenbinary}
\end{footnotesize}
\end{center}
\end{adjustwidth}
\end{minipage}
\end{table}
\setlength{\tabcolsep}{6pt}

\subsubsection{Inverse probability weighting with a misspecified weighting model}
Inverse probability weighting can yield biased estimates if the model for computing the weights is misspecified \cite{seaman2011}. Nevertheless, for the simulations in this paper, selection depended on the risk factor and the confounder but a reasonable approximation to the true causal effect was obtained via weighting by the risk factor only.

The behaviour of inverse probability weighting can be significantly worse if the confounder only has a weak influence on the risk factor. We illustrate this by conducting a simulation similar to that of Table~\ref{scentrim}, with a weak confounder effect on the risk factor. We set $\gamma_U = 1$ and  $\alpha_U = \sqrt{0.1}$ and leave the other parameters unchanged. Results are presented in Supplementary Table~\ref{sceniptw}.

In this simulation, causal effect estimates are subject to significant bias when the risk factor--selection effect is strong. This is the case even when using the inverse probability weighting approach. Again, trimming weights was of little consequence in this example.

\setlength{\tabcolsep}{4pt}
\begin{table}[htbp]
\begin{minipage}{\textwidth}
\begin{adjustwidth}{-0.4cm}{-0.4cm}
\begin{center}
\begin{footnotesize}
\centering
\begin{tabular}[c]{c|cccc|cccc|cccc}
\hline
$\gamma_U = 1$	&  Med   &   SD    &  Med SE  &  Type 1 &  Med   &   SD    &  Med SE  &  Type 1 &  Med   &   SD    &  Med SE  &  Type 1   \\
\hline
$\gamma_X$        & \multicolumn{4}{c|}{No trimming}        & \multicolumn{4}{c|}{Trimming at 99\%}   & \multicolumn{4}{c}{Trimming at 95\%}      \\
\hline
$-2$        &  0.158 & 0.112 & 0.080 & 51.1 \%	&  0.134 & 0.105 & 0.088 & 37.3 \%	&  0.108 & 0.106 & 0.097 & 22.9 \% \\
$-1$        &  0.101 & 0.074 & 0.072 & 29.3 \%	&  0.096 & 0.075 & 0.074 & 26.2 \%	&  0.088 & 0.078 & 0.076 & 21.7 \% \\
$-0.5$      &  0.053 & 0.069 & 0.069 & 12.5 \%	&  0.053 & 0.069 & 0.069 & 12.2 \%	&  0.051 & 0.070 & 0.070 & 11.8 \% \\
$-0.2$      &  0.024 & 0.068 & 0.068 &  6.7 \%	&  0.024 & 0.068 & 0.068 &  6.7 \%	&  0.023 & 0.068 & 0.068 &  6.5 \% \\
$0$         &  0.003 & 0.067 & 0.067 &  5.3 \%	&  0.002 & 0.068 & 0.067 &  5.2 \%	& -0.001 & 0.069 & 0.068 &  5.1 \% \\
$0.2$       & -0.016 & 0.068 & 0.065 &  6.6 \%	& -0.019 & 0.069 & 0.066 &  6.8 \%	& -0.024 & 0.071 & 0.068 &  7.3 \% \\
$0.5$       & -0.045 & 0.071 & 0.065 & 13.0 \%	& -0.050 & 0.072 & 0.067 & 13.9 \%	& -0.060 & 0.075 & 0.070 & 15.3 \% \\
$1$         & -0.086 & 0.080 & 0.064 & 31.4 \%	& -0.101 & 0.080 & 0.068 & 33.6 \%	& -0.118 & 0.083 & 0.074 & 36.0 \% \\
$2$         & -0.158 & 1.325 & 0.066 & 61.3 \%	& -0.192 & 0.107 & 0.077 & 65.5 \%	& -0.220 & 0.105 & 0.088 & 68.2 \% \\
\hline
\end{tabular}
\caption{Median, standard deviation (SD), median standard error (med SE) of estimates and empirical Type 1 error rate (\%) for risk factor-outcome causal associations with a misspecified inverse probability weighting model ($\gamma_U = 1$) and a weak confounder--risk factor effect ($\alpha_U = \sqrt{0.1}$), for different values of the selection effect ($\gamma_X$).} \label{sceniptw}
\end{footnotesize} %
\end{center}
\end{adjustwidth}
\end{minipage}
\end{table}
\setlength{\tabcolsep}{6pt}

\end{document}